\documentclass[]{aa}

\usepackage{graphicx}
\usepackage{caption}
\usepackage{txfonts,color}

\begin{document}

\title{Abundance ratios in the hot ISM of elliptical galaxies}

\author {A. Pipino\inst{1,2} \thanks{email to: pipinoa@phys.ethz.ch, pipino@oats.inaf.it} \and F. Matteucci\inst{1,3} }
 \institute{
  Dipartimento di Astronomia, Universit\`a di Trieste, via
  G.B. Tiepolo 11, I-34131, Trieste, Italy \and
  Institut fur Astronomie, ETH Zurich, 8093 Zurich, CH,
\and I.N.A.F. Osservatorio Astronomico di
  Trieste, via G.B. Tiepolo 11, I-34131, Trieste, Italy }

\date{Received xxxx / Accepted xxxx}

 \abstract{}{To constrain the recipes
put forth to both solve the \emph{theoretical}
Fe discrepancy in the hot interstellar medium of elliptical galaxies and \emph{at the same time} explain the
 [$\alpha$/Fe] ratios. }{In order to do so we use the latest theoretical nucleosynthetic yields,
we incorporate the dust, we explore differing SNIa progenitor scenarios
by means of a self-consistent chemical evolution model which reproduces
the properties of the stellar populations in elliptical galaxies.}{Models with  Fe-only dust 
and/or an \emph{effective} SNIa rate  
achieve a better agreement with the observed Fe abundance. 
However, a suitable modification to the SNIa yield with respect to
the standard W7 model is needed to fully match the abundance ratio pattern. 
The 2D explosion model C-DDT by Maeda et al. (2010) is a promising candidate for reproducing
the [Fe/H] and the [$\alpha$/Fe] ratios.}{}

\keywords{ISM: dust, extinction; ISM: abundances; galaxies: ellipticals: chemical abundances, formation and evolution }

\titlerunning{Abundance rations in the ISM}
\authorrunning{Pipino \& Matteucci}
\maketitle

\section{Introduction}

The spectra of the hot X-Ray emitting interstellar medium (ISM)
of elliptical galaxies carry valuable information on the composition
of both the stellar and the Type Ia supernova (SNIa) ejecta in the
late phases of the galactic evolution.
Recent \emph{Chandra} and \emph{XMM} measurements (e.g. Buote, 2002,
Xu et al., 2002, Gastaldello \& Molendi, 2002, Buote et al., 2003, 
Kim \& Fabbiano, 2004, Humphrey \& Buote, 2006, Tawara et al., 2008,
Ji et al., 2009, Loewenstein \& Davis, 2010) agree that X-ray bright
elliptical galaxies exhibit [Fe/H]$\sim 0 - 0.3$ dex,  namely typically solar
or twice. Overall, one can estimate that the contribution of SNIa to the metal enrichment
of the hot ISM is quite important, being probably more than 60\% (e.g. Humphrey
\& Buote, 2006).
Such an Fe abundance is similar to the average
abundance in the stars of the same galaxies, as inferred from absorption lines
in optical stellar spectra (Humphrey
\& Buote, 2006), whose observed stellar 
pattern 9e.g. Graves et al., 2007) is well reproduced by the most recent monolithic formation models (e.g. Pipino
et al., 2009a).

However, since the earlier theoretical attempts by Loewenstein \& Mathews (1991, see also Renzini et al., 1993, Pipino et al., 2005),
all the monolithic collapse/galactic wind models agree in predicting quite high abundances, a factor
of 2-3 higher than the observational measurements.
Such a high abundance is a clear consequence of the simple assumptions made in the model, 
where all the iron produced by SNIa after the star formation and wind phases, immediately mixes with the ISM.
We will refer to this as the \emph{theoretical Fe discrepancy}.
Therefore, either some modifications to the simple galactic wind scheme are required
or a completely different galaxy formation scenario should be adopted. 
The possible physical mechanisms invoked to solve the  discrepancy between the predicted Fe abundances and the observed ones (see
also Renzini et al., 1991, Arimoto et al.\ 1997 for earlier studies), can be 
roughly grouped in several classes: differential ejection of Fe; 
iron hidden in a colder phase (warm gas or dust); a significant dilution from freshly accreted gas
with primordial abundances; modifications to the SNIa rate. Therefore, some other physical processes must be added to
the basic wind model.
For instance, very recent simulations by Tang \& Wang (2010) show that the hot and low-density SNIa bubbles
move outward faster than the surrounding medium. Such a differential motion reduces the [Fe/H] by more than 0.2 dex 
with respect to the case in which the Fe is immediately mixed with the ambient gas.
Moreover, because of the high temperature and low density inside the bubbles, the contribution 
of such Fe-rich gas to the total X-ray emission is negligible, hence the ``emission weighted'' abundance
that we infer from observations mirrors the metallicity of the dense (and relatively less Fe enriched)
shells swept by the SNIa remnants.

Mathews \& Brighenti (2003) suggested that a sizable
fraction of metals never enters in the hot phase due to dust-assisted cooling. 
The problem with the cooling scenario is that the observed lines at $\sim$ $10^6$K strongly constrain the amount
of metals that can cool (e.g. Peterson et al., 2001, Xu et al., 2002). 
However, warm ISM is observed in ellipticals and its metallicity is nearly solar (e.g. Athey
\& Bregman, 2009), hence consistent with the measurements in both stars and hot ISM.
According to Brighenti \& Mathews (2005), Fe-rich ejecta may cool faster than the ambient medium
because of its large radiative emissivity. 

Finally, several authors (e.g. Renzini et al., 1993, Arimoto et al., 1997, Brighenti \& Matthews, 2005, Loewenstein \& Davis, 2010)
claim that the actual SNIa rate in ellipticals is at most 1/4 of the observationally derived one,
but no study offered a clear explanation of the reason.
A natural consequence of this assumption is that the [Fe/H] is lowered.
In principle, elliptical galaxies have slightly different star formation histories, hence
slightly different SNIa rate. Those with a higher average SNIa rate in the recent
past may have undergone a galactic wind till later times and perhaps
they have not yet built enough gas mass to be detectable in the X-rays.
There are some recent suggestions indicating that the present-day SNIa rate correlates 
with the colour of the galaxy (Mannucci et al., 2005). 
Also, different progenitor systems
give rise to different SNIa rates. In fact, even if the total number of SNIa ever exploded and the
present-day SNIa rate are the same, the distribution in time of the SNIa events
changes. This affects the fraction of Fe ejected in the wind
as opposed to the Fe trapped in the hot ISM. 

As far as the abundance ratios are concerned,
observations (e.g. Humphrey \& Buote, 2006, Ji et al., 2009, Loewenstein \& Davis, 2010)
generally agree that [Si/Fe] is solar to super-solar, [Mg/Fe]$\sim 0$ and that
the O is depleted with respect to Mg.
While a general consensus seem to emerge from comparing \emph{Chandra} and \emph{XMM} data,
a caveat is that different abundance patterns may be inferred from different instruments
on board the same telescope, the largest variations being for the O abundance (Ji et al., 2009).
According to Loewenstein \& Davis (2010), { who also measure a strongly super-solar [Ca/Fe],} neither a single model 
nor a single set of SNIa yields can reproduce the entire abundance pattern
observed in NGC 4472.
Moreover, the theoretical Fe discrepancy affects the [element/Fe] ratios too,
in the sense that the predictions would exhibit - in general - lower abundance ratios than the
observations, unless a lower than observed effective SNIa rate is adopted.
 Unfortunately, most of the models previously discussed
to address the \emph{theoretical} Fe discrepancy do not track the evolution of species other than Fe, therefore
they cannot be thoroughly tested against the ISM abundance pattern. 
Also, it is worth
reminding that most of the galactic wind models aimed at reproducing the hot ISM do not study the formation of the galaxy, nor do they
make predictions on the stellar abundances. Indeed, in many cases the simulation starts after the galaxy
has been already assembled.

In our previous work we have already tested a mild dilution from pristine gas (Pipino et al., 2005),
or dust (Calura et al., 2008), to show that the predicted [Fe/H] can be reduced
so that the disagreement between theory and observation decreases.
Here, we revisit the issue by means of updated dust prescriptions 
and contrast the predicted element ratios [X/Fe] to the observed
ones. Moreover, the present work is aimed at constraining some of the above 
mentioned theoretical solution
put forth to address the theoretical
Fe discrepancy. \emph{At the same time}, we aim at explaining the [$\alpha$/Fe] ratios 
by means of a self-consistent chemical evolution model which reproduces
the properties of the stellar populations in elliptical galaxies.
In order to do so we use the latest theoretical nucleosynthetic yields,
incorporate the dust in the model and explore different SNIa progenitor 
scenarios.

The basic features of our chemical evolution scheme  as well
as the model set-up are summarized in Sec.~\ref{thecode}.
The results are presented and discussed in Secs.~\ref{fe} and~\ref{abu}.
Conclusions are drawn in Sec.~\ref{concl}.

{ The adopted solar abundances are
those by Grevesse \& Sauval (1998).}

\section{The chemical evolution model for elliptical galaxies}
\label{thecode}

\subsection{The chemical evolution model}
We use the model presented in Pipino et al. (2005 - see also Calura et al., 2008
and Pipino et al., 2011 for the formulation of dust evolution).
The galaxy evolves as an open box in which the initial gas mass, with primordial chemical composition,
rapidly  collapses, in a time scale $\tau$,
into the potential well of a dark matter halo.

The rapid collapse triggers an intense and rapid  star formation process, which can be considered as a
starburst lasting until a galactic  wind,
powered by the thermal energy injected by stellar winds and SN
explosions, occurs. At that time ($t_{gw}$, hereafter), the thermal energy  equals the
binding energy of gas. After $t_{gw}$, the galaxies evolve passively.
In the particular model that we present, $t_{gw}=1.2$Gyr. The wind stops at $t_{stop}=11$Gyr.

The equation of chemical evolution for the element $i$ takes the following form:
\begin{eqnarray}\label{main}
{d G_i (t) \over d t}  &= & -\psi (t) X_i (t)\,   \nonumber \\
& & +\int_{M_L}^{M_{Bm}} \psi (t-\tau_m) Q_{\rm mi}(t-\tau_m) \phi (m) dm\,  \nonumber \\
& & + A\int_{M_{Bm}}^{M_{BM}} \phi (m) \nonumber  \\
& &\cdot \left [ \int_{\rm\mu_{\rm min}}^{0.5} f(\mu) Q_{\rm mi}(t-\tau_{\rm m_2}) \psi (t-\tau_{\rm m_2}) d\mu \right ] dm \,  \nonumber \\
& &  +(1-A)\, \int_{\rm M_{\rm Bm}}^{M_{\rm BM}} \psi (t-\tau_m) Q_{\rm mi}(t-\tau_m) \phi (m) dm\,  \nonumber \\
& &  +\int_{\rm M_{\rm BM}}^{M_U} \psi (t-\tau_m) Q_{\rm mi}(t-\tau_m) \phi (m) dm\,  \nonumber \\
& & +({d G_i (t) \over d t})_{\rm infall}
-W(t)X_i (t)
+({d G_i (t) \over d t})_{\rm acc}
\, ,
\end{eqnarray}
where $G_i (t)$
is {the  fractional mass  of the element \emph{i} at the time \emph{t} in the ISM. $X_i (t)$
is defined as the abundance by mass of the element \emph{i}.
By definition $\sum_i X_i=1$.

The first term of Eq. \ref{main} gives the rate at which the element
$i$ is  subtracted from the ISM by the star formation process.  The variable $\psi$ is
the star formation rate calculated according to the following law:
\begin{equation}
\psi (t)= \nu\cdot G (t),
\end{equation}
where $G(t)$ is the fractional mass of gas and
$\nu$ is a constant which represents the star formation efficiency. 

The second term is the rate at which each element is restored into
the ISM by single stars with masses in the range $M_{L}$ -
$M_{Bm}$, where $M_{L}$ is the minimum mass contributing, at a
given time $t$, to chemical enrichment (the minimum is $0.8
M_{\odot}$) and $M_{Bm}$ is the minimum binary mass allowed for
binary systems giving rise to SNIa.  The initial mass function (IMF) has the Salpeter(1955)
form $\phi (m)\propto m^{-(1+1.35)}$,  and it is
normalized to unity in the mass interval $0.1 - 40 M_{\odot}$. In
particular, $Q_{mi}(t-\tau_m)=Q_{mi} \, X(t-\tau_m)$, where $Q_{mi}$
is a matrix which calculates for any star of a given mass $m$ the
amount of the newly processed and already present element $i$, which
is returned to the ISM. The quantity $\tau_m$ is the lifetime of a
star of mass $m$ (Padovani \& Matteucci, 1993).

The third term represents the enrichment by SN Ia. {
In the following we will assume the Single Degenerate (SD) scenario for our fiducial model:
a C-O white dwarf plus a red giant (Whelan \& Iben 1973).  Following  Greggio $\&$ Renzini (1983) and Matteucci $\&$ Greggio (1986), the SNIa rate is:
\begin{equation}
R_{SNIa}=A\int^{M_{BM}}_{M_{Bm}}{\phi(M_B) \int^{0.5}_{\mu_min}{f(\mu)
\psi(t- \tau_{M_{2}})d \mu \, dM_{B}}}\, ,
\label{snia}
\end{equation}
where $M_{\rm B}$ is the total mass of the
binary system, $M_{Bm}=3 M_{\odot}$ and $M_{BM}=16 M_{\odot}$ are the minimum
and maximum masses allowed for the adopted progenitor systems, respectively.
$\mu=M_2/M_{\rm B}$ is the mass fraction of the secondary, which
is assumed to follow the distribution law:
\begin{equation}
f(\mu)=2^{1+\gamma}(1+\gamma)\mu^\gamma\, 
\end{equation} and $\gamma=2$.
Finally, $\mu_{min}$ is the minimum mass fraction contributing to the SNIa rate at the time $t$, and is given by:\\
\begin{equation}
\mu_{min}=max \left \{ \frac{M_{2}}{M_{B}}, \frac{M_{2}-0.5M_{B}}{M_{B}} \right \}
\end{equation} 
The predicted type Ia SN explosion rate is constrained
to reproduce the present day observed value (Mannucci et al., 2008), by
fixing the parameter $A$ in Eq. (\ref{main}). In particular, $A$ represents
the fraction of binary systems in the IMF which are
able to give rise to SNIa explosions. The fiducial model has $A=0.09$.
We present also a model where we force a sudden transition from $A=0.09$ to 
$A<0.09$ after a suitable time ($> t_{gw}$),
in order to model a reduced \emph{effective} SNIa rate in the late evolutionary phases.

We will also present a model which considers the Double Degenerate (DD) \emph{wide} scenario
(see Greggio, 2005 and Valiante et al., 2009 for details). We note that when such a scenario is constrained
to reproduce the present day SNIa rate, the total number of SNIa (hence
the total mass of synthesized Fe) will be lower than that in the fiducial case by a factor of 4. 
This happens because in the DD scenario there is a further delay introduced
by the timescale of the merging of the two white dwarfs with respect to the SD
scenario. In practice, many more systems did not have time to explode in a Hubble time.
Also, we tested the empirical progenitor model by Mannucci et al. (2005, see also Matteucci et al., 2006).
Given the diversity in SNIa light curves, one cannot exclude that different progenitors
and explosion characteristics coexist.}

The fourth term { gives the contribution of either single stars (including SNII) or binaries not giving rise to SNIa in the mass range $M_{Bm} - M_{BM}$.} The fifth term represent the enrichment by stars more massive than $M_{BM}$ 
The initial galactic gas infall phase enters the equation via the sixth
term, for which adopt the formula:
\begin{equation}
({d G_i (t) \over d t})_{infall}= X_{i,infall} C e^{-{t \over
\tau}}\, ,
\end{equation}
where $X_{i,infall}$ describes the chemical composition of the
accreted gas, assumed to be primordial. $C$ is a constant obtained
by integrating the infall law over time.
Finally, $\tau$ is the infall timescale.

The last two terms account for the mass ejection during the galactic
wind and possible late-time accretion of surround pristine gas.

{  The predictions will be presented as averaged over 30 kpc. This choice
is consistent with the inner region probed in Loewenstein \& Davis (2010) with Suzaku
in the specific case of NGC 4472, which will be the galaxy that we use as the prototypical observational case,
although the model has not been specifically tailored for this galaxy.
It is worth mentioning that for this galaxy no particular ISM abundance gradients are reported (Loewenstein \& Davis, 2010),
but for the [Mg/Fe] which roughly decreases by a half in the outer regions. 
Therefore we do not expect unaccounted gradients to affect our specific comparison. 
The study of metallicity gradients in the hot ISM will be the topic of future work.}

\subsection{Stellar yields}

The yields used in this paper are as follows:
\begin{enumerate}
\item
For single low and intermediate mass stars ($0.8 \le M/M_{\odot} \le
8$) we make use of the yields by van den Hoek \& Groenewegen (1997)  as a function of metallicity.
\item For SNeIa we adopt either the W7 or the WDD1 models from standard 1D simulations
(Iwamoto et al., 1999) as well as the newer C-DDT model from 2D explosion simulation (Maeda et al., 2010).
In particular, the model with deflagration to detonation transition (DDT)
seem to better reproduce the layered structure of the ejecta (Maeda et al., 2010
and references therein) with respect to the standard deflagration model W7.
We also considered a case in which we adopt the recently suggested empirical yields
by Fran\c cois et al. (2004). These yields are a revised version of the Woosley
\& Weaver (1995, for SNII) and Iwamoto et al. (1999, for SNIa) calculations adjusted
to best fit the chemical abundances in the Milky Way. In our particular case, in which
we focus on the late time chemical evolution, we simply
applied the changes to the SNIa yields. According to Fran\c cois et al. (2004)  the Mg production in SNIa must increase by a factor of 5 relative to standard yields,
whereas the O and Fe production should not change (models labelled as \emph{W7,5xMg} hereafter).

\item For SNeII we use the yields by Kobayashi et al. (2006) for the solar composition
as the fiducial case (K002, hereafter). We also ran a model with the Kobayashi
et al. (2006) Z=0.001 case (K0001, hereafter).

\end{enumerate}

\begin{table*}
\begin{minipage}{120mm}
\scriptsize
\begin{flushleft}
\caption{Model set-up and average stellar abundance ratios}
\begin{tabular}{l|lc|lc|l|lllll}
\hline
Model&        &  YIELDS&          &SNIa RATE &  DUST  & [$<$Fe/H$>$]&[$<$Mg/Fe$>$]&[$<$Ca/Mg$>$]&[$<$Ca/Fe$>$]&[$<\alpha$/Fe$>$]\\
name & SNII   &  SNIa  &  progen. & A        &        \\   
\hline

1a   & K002   & W7     &   SD     & 0.09     &  off   & 0.133   & 0.209     & -0.07    &  0.038  & 0.308\\
1ad  & K002   & W7     &   SD     & 0.09     &  on    \\
1afd & K002   & W7     &   SD     & 0.09     &  only Fe   \\

1b   & K0001    & W7     &   SD     & 0.09     &  off   \\

1c   & K002   & W7,5xMg &   SD     & 0.09     &  off   \\
1cd  & K002   & W7,5xMg &   SD     & 0.09     &  on   \\
1cfd & K002   & W7,5xMg &   SD     & 0.09     &  only Fe   \\


\hline

2a   & K002   & WDD1   &   SD     & 0.09     &  off   & 0.034 & 0.251  & 0.045  & 0.161 & 0.354\\
2ad  & K002   & WDD1   &   SD     & 0.09     &  on   \\
2afd & K002   & WDD1   &   SD     & 0.09     &  only Fe   \\


\hline

3a   & K002   & C-DDT     &   SD     & 0.09     &  off   & -0.08 & 0.329  & -0.188 & 0.100 & 0.416\\
3ad  & K002   & C-DDT     &   SD     & 0.09     &  on   \\
3afd & K002   & C-DDT     &   SD     & 0.09     &  only Fe   \\

3b   & K0001   & C-DDT     &   SD     & 0.09     &  off   \\


\hline

4a   & K002   & W7     &   SD     & 0.09$t<t_{gw}$   &  off   & 0.133   & 0.209     & -0.07    &  0.038  & 0.308\\
     &        &        &          & 0.01 otherwise   &        \\

4b   & K002   & W7    &   SD     & 0.09$t<11Gyr$     &  off   \\
     &        &        &          & 0.01 otherwise   &        \\

4c   & K002   & W7      &   SD     & 0.09$t<8Gyr$     &  off   \\
     &        &        &          & 0.01 otherwise   &        \\

     &        &        &          & 0.01 otherwise   &        \\

\hline

5a   & K002   & W7     &   DD     & $f(t)$     &  off  & -0.348 & 0.421   & -0.175   & 0.212  & 0.514 \\

\hline


\hline
\end{tabular}
The reader should note that once the yields and the SNIa progenitor are specified, the abundances in the stars
do not change. Therefore we report them for each ``class'' of models, irrespective of the $t>t_{gw}$ modifications.
\label{setup}
\end{flushleft}
\end{minipage}
\end{table*}

\begin{figure}
\begin{center}
\includegraphics[height=8cm,width=8cm]{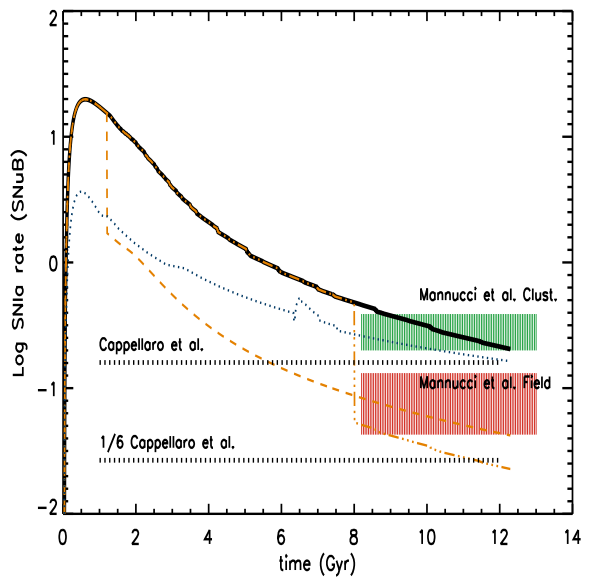} 
\caption{SNIa rate (in SNuB) versus time predicted by models 1a (solid), 4a (dashed), 4c (dot-dashed) and 5a 
(dot-dotted), respectively. The shaded areas give the observed range in clusters and lower
density environments by Mannucci et al. (2008), respectively. The horizontal lines mark the present
day rate by Cappellaro et al. (1998), that has been used by many previous works and a rate that
is 1/6 of the Cappellaro et al. value, as adopted by Loewenstein \& Davis (2010). \label{fig0}}
\end{center}
\end{figure}

\subsection{Dust model}

We refer the reader to Calura et al. (2008) and Pipino et al. (2011) for
the details of the chemical evolution model with dust, constrained by 
both low- and high-redshift observables.
Here we recall the main assumptions. Only the main refractory elements, 
C, O, Mg, Si, S, Ca, Fe, deplete into dust,
and we assume that stars can produce two different types of grains: i) silicate dust, 
composed by O, Mg, Si, S, Ca and Fe, and ii) carbon dust, composed by C.  
Following Dwek (1998), we consider 
that the dust producers are  low and intermediate mass stars,  Type Ia SNe and  Type II SNe.
 
The dust destruction is assumed to occur  in SN shocks.
Following McKee (1989), the SN shock term adopts different destruction efficiencies before and after the galactic wind. 

We assume the dust accretion occurs only during the starburst epoch in dense molecular clouds, where volatile elements can condensate onto pre-existing grain cores, originating a volatile part called mantle (Dwek, 1998). Therefore, after the galactic wind occurs and star formation stops there is only dust production from long living stars.


\subsection{The models}

The comprehensive list of all the models is presented in Table~\ref{setup}.
In the same table we show the predicted average stellar abundance ratios (see Pipino et al., 2006, 2008 for their
definitions and related discussion).
The reader should note that once the yields and the SNIa progenitors are specified, the abundances in the stars
do not change. Therefore we report them only once for each ``class'' of models, irrespective of the $t>t_{gw}$ evolution.
The predicted abundance and abundance ratios agree very well with recent data by Graves et al. (2007).
In particular, in models of the classes ``1'' and ``4'' - which
basically correspond to the models presented in Pipino \& Matteucci (2004) and Pipino 
et al. (2009a) - the [Fe/H] is found to be mildly over-solar, the Mg (as well as the $\alpha$ elements as a whole) is enhanced, 
while the Ca tracks more closely Fe rather than other $\alpha$ elements, since it is produced by SNeIa in a non-negligible 
way (see Pipino et al., 2009a). Model 5a has very few SNIa at $t<t_{gw}$, hence a very high $\alpha$-enhancement, even
higher than the observations. Models of the classes ``2'' and ``3'' are in between these
two extremes and in broad agreement with observations (e.g. Graves et al., 2007).
{ Model 2 predictions are also those which best match the observationally inferred stellar Fe and Mg abundances (Thomas
et al., 2005, Humphrey \& Buote, 2006) for the specific case of NGC 4472.}
Unfortunately, we cannot derive the O, Si, and Ni abundances in integrated
stellar spectra, therefore the models are unconstrained on this side. { We predict that the stellar
[O/Mg]= +0.1 dex. This is naturally expected from the specific choice
of the Kobayashi et al. yields. The effect of a change in the SNIa progenitor is at the 0.02 dex level.}
The SNIa rate versus time is shown
in Fig.~\ref{fig0}. In particular, we show the rate predicted by models 1a (solid), 4a (dashed), 4c (dot-dashed) and 5a 
(dot-dot-dashed), respectively. The rate predicted by model 1a is exactly the same
for all the other models grouped under the categories \emph{1,2, and 3}. 
Model 5a exhibits a slower late time decrease in the SNIa with respect to the fiducial case.
The shaded area gives the observed range in clusters and lower
density environments (``field'') by Mannucci et al. (2008). The horizontal lines mark an older estimate
of the present day rate (Cappellaro et al., 1998), that has been used by many previous works and a rate that
is 1/6 of the Cappellaro et al. value, as an example of
an \emph{effective}\footnote{Namely the value that best reproduces the Fe abundance
as opposed to the actual observed SNIa rate} SNIa rate adopted by other authors (e.g. Loewenstein \& Davis, 2010). 

{ Before discussing the metal abundances it is important to stress that these
models predict a final ISM mass comparable to the observed one.
Pipino et al. (2005) selected the models
by requiring that they reproduced the gas/star mass ratio and the $L_X$ at that given mass in agreement with observations (and
the large observational spread).
The models presented in the submitted paper basically differ from those of Pipino et al. (2005) only
for the nucleosynthesis, hence we expect that the features other than the abundances and abundance ratios to be preserved.
The current model predicts no outflow in the final Gyr of evolution, and a final gas mass
of $\sim 2\times 10^9 M_{\odot}$. Although the model was not tailored on any specific galaxy, it is useful
to compare this value to the reported ISM masses for NGC 4472, which range from
$M_{gas} (< 13 kpc) \sim 10^9 M_{\odot}$ (Athey, 2007, Chandra data) to $M_{gas}(<30 kpc) \sim 5\times 10^9$ 
(Irwin \& Sarazin, 1996, with older ROSAT data).}

%
%

\section{The Fe abundance in the ISM}
\label{fe}

Concerning the study of the chemical composition of the ISM, we
first analyze the iron abundance, which is the best determined
among all the metals.

The natural predictions in a monolithic collapse + galactic wind model for the Fe abundance
are shown in Fig.~\ref{fig1} (solid lines).
{ The model predictions are qualitatively compared to observations in Fig.~\ref{fig1}. In particular the typical range in the
observed values (shaded areas, c.f. Introduction)
are given for reference.}
The predicted high abundance is a clear consequence of the simplest possible model, 
where all the iron produced
after the wind has stopped is retained by the galaxies, whereas most of the H
present during the star forming phase has been ejected and only a minor fraction
is restored by the dying stars ($<$1.5-2 M$_{\odot}$). The { Fe pollution in the gas
is dominated by SNIa ejecta}.

At z$\sim$0 these models
predict a very high ($>$0.5 dex) [Fe/H], as expected from the continuous Fe production by SNIa.
This is at variance with observations (shaded areas in Fig.~\ref{fig1}). 
However, as opposed to previous models where the predicted Fe abundances were offset from the observed
values by a factor of 3 or more, the latest model predictions are only within a factor of 2 of
the observations. One reason is the fact that updated versions of the SNIa W7 model produce
$\sim0.6 M_{\odot}$ of Fe instead of the classically used mass of $\sim0.74 M_{\odot}$. This
change by itself accounts for a 0.1 dex decrease in the predicted [Fe/H]. 
Also, models differing in the the duration of the wind and its
mass-loading, will have slightly different final [Fe/H] in the gas. 

The [Fe/H] in the hot ISM can also be used as an age indicator, because of
its strong evolution with time (Fig.~\ref{fig1}). [Fe/H]$>$1 are typical
of ejecta that do not mix with the ambient medium, and also the likely composition of the wind
in its intermediate to late phases, namely when most of the pristine H in the galaxy left over
after the star formation process has already been expelled.
At very late times, when the hot medium has been built-up, the Fe abundance decreases because
of both the dilution by dying low mass stars and the decline in the SNIa rate.
Therefore, galaxies with similar stellar mass but different [Fe/H] and $L_X$ (that roughly tracks
the hot gas mass) may be just in different stages of the halo build-up.

Also, we note that
the observed SNIa present-day explosion rate is shown to have little dependence
on the galactic mass (e.g. Mannucci et al., 2008), hence on the luminosity in the optical bands. Therefore,
one can naively expect a mild (if any) correlation between the Fe abundance in the 
hot ISM and the galaxy mass. That is, while the stellar abundances obey to a very tight galactic mass-metallicity relationship
(e.g. O'Connel, 1976), this is not true as far as the hot gas abundances are concerned.
This is indeed what is observed by Humphrey \& Buote (2006) and what is predicted by the wind models (Pipino et al., 2005).

\subsection{Solutions to the theoretical Fe discrepancy in wind models}
\label{solutions}

As discussed in the Introduction, possible physical mechanisms often invoked to solve the  discrepancy between the predicted
(theoretical) Fe abundances and the observed one (see
also Renzini et al., 1991, Arimoto et al.\ 1997), can be roughly grouped in several classes: differential ejection of Fe
hidden in a colder phase (warm gas or dust); a significant dilution from freshly accreted gas
with primordial abundances; the modifications to the SNIa rate (the SNIa progenitor/explosion scenario). \\

We do not attempt to test here all the proposed scenarios. Moreover, some of them can be safely discussed
only by means of hydrodynamical models.
Here, we will focus on some of them. Nonetheless, we believe that the result will be rather general.
In particular, we will focus on
models with dust (marked with \emph{d} in Table~\ref{setup}),
models with SNIa nucleosynthesis differing from the standard W7 explosion (model 1c and categories \emph{2, 3} in Table~\ref{setup}),
models with modified SNIa rates (categories \emph{4, 5} in Table~\ref{setup}). These
latter models either mimic differential ejection (or cooling) of SNIa ejecta (\emph{4}) or
a different progenitor scenario (\emph{5}).

\begin{figure}
\begin{center}
\includegraphics[height=10cm,width=9cm]{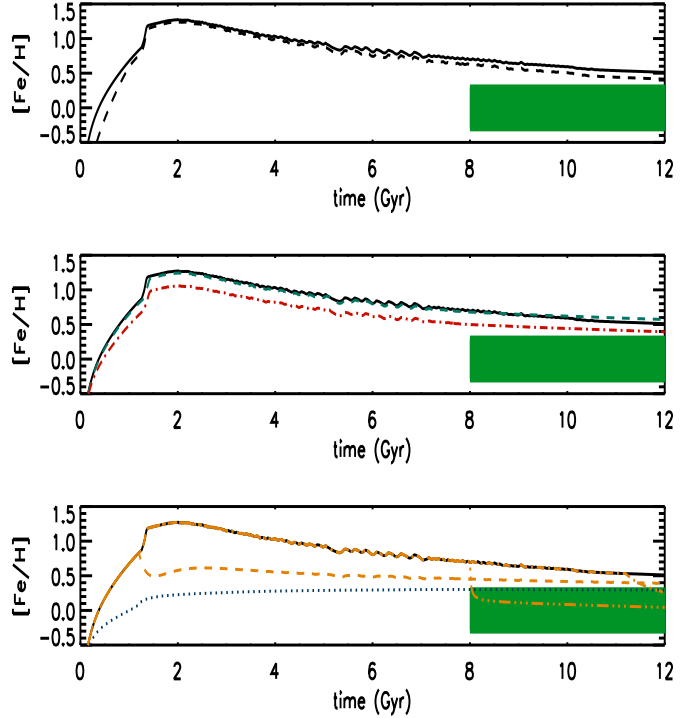}
\caption[]{Lines: Predicted [Fe/H] in the ISM versus time. Shaded region: { range of typical observed values from the literature
(errors not included)}. Upper panel: Model 1a (fiducial, solid), Model 1ad (fiducial+dust, dashed) .
Middle panel: Model 1a (fiducial, solid), Model 2a (WDD1 SNIa, dashed), Model 3a (C-DTT SNIa, dot-dashed).
Lower panel: Model 1a (fiducial, solid) compared to Models 4a (dashed), Model 4b (dotted), and Model 5a
(dot-dashed) with differing SNIa rate. \label{fig1}}
\end{center}
\end{figure}

\clearpage

\begin{figure}
\begin{center}
\includegraphics[height=18cm,width=22cm,angle=90]{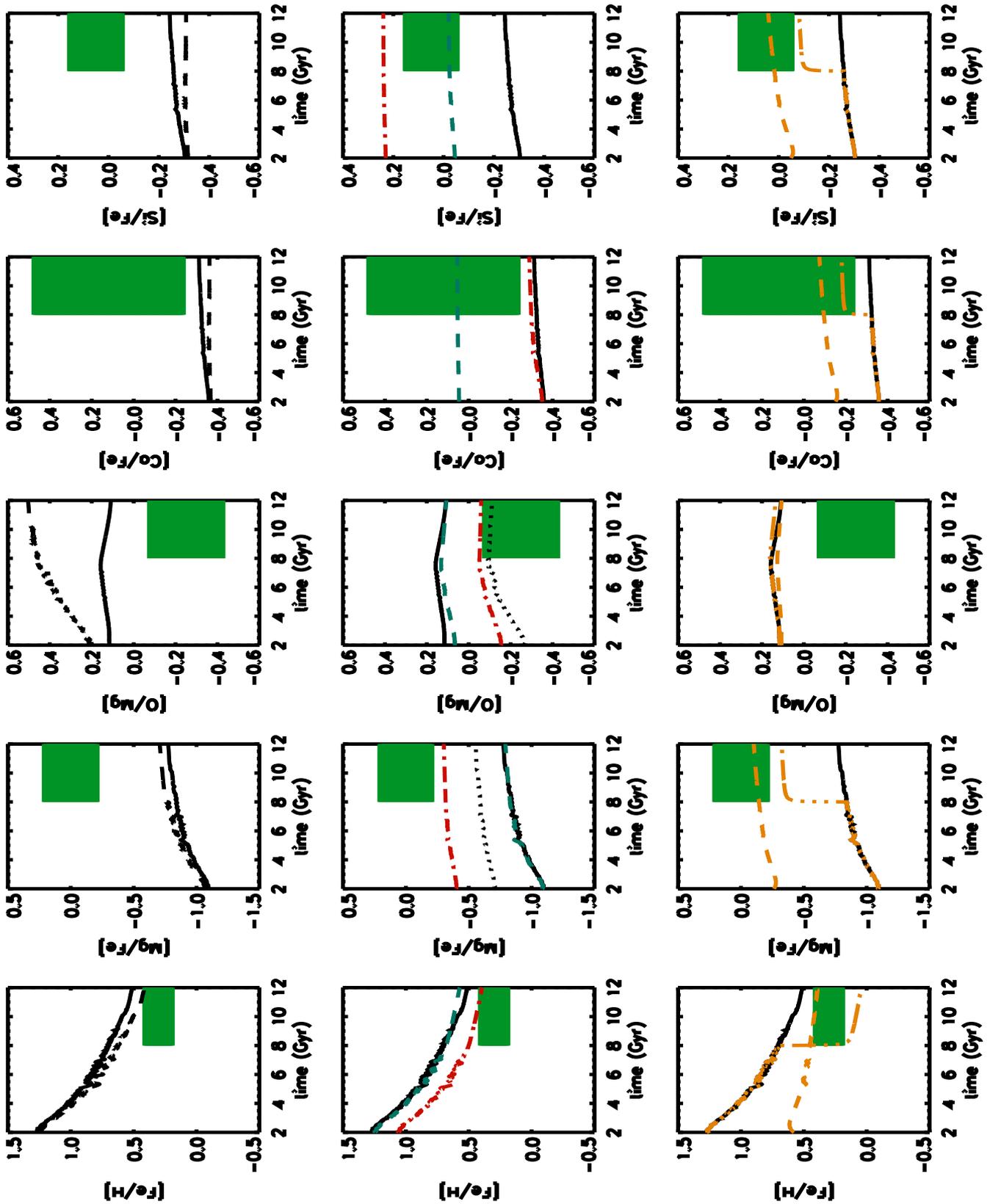}
\caption[width=18cm]{\scriptsize [$\alpha$/Fe] abundance ratios versus time. The shaded areas represent
the observed values and the 99.6\% uncertainties in NGC 4472 from Loewenstein \& Davis (2010). \emph{Results
for [Mg/Fe]} Upper row: fiducial model 1a (solid) compared to model 1afd (dashed).
Middle row: fiducial model 1a (solid) compared to models 1c (dotted), 2a (dashed), 3a (dot-dashed), respectively.
Lower row: fiducial model 1a (solid) compared to models 4a (dot-dashed) and 4c (dashed),
respectively. \emph{Results
for [O/Mg], [Ca/Fe] and [Si/Fe]} Models are coded as in the [Mg/Fe] column except for the model 1afd and in the upper panel
that is replaced by models 1ad. { The [Fe/H] evolution from Fig.~\ref{fig1} is also reported for comparison to the
specific Fe abundance measured in NGC 4472}. \label{fig2}} 
\end{center} 
\end{figure}

\clearpage

Pipino (2011, see also Calura et al., 2008) showed the predicted  evolution of the Fe abundance in the hot phase of a giant elliptical 
galaxy  with dust.
They assumed that no sputtering\footnote{According to Itoh (1989),
the density and temperature of the hot ISM in ellipticals are such that
the lifetime of a Fe dust grain can be as short as $\sim 10^{6-7} yr$. Therefore Arimoto
et al. (1997) argued that it is unlikely to store a large amount of Fe in dust.
Such an argument was supported by the relatively small mass of dust that could be observed
in lanes and clouds.
Recent observations, however, claim that the dust mass in ellipticals could be as high as $\sim 10^{6-7}M_{\odot}$,
(e.g. Temi et al.\ 2004).} from the hot ISM is occurring and that as much
as 80\% of SNIa ejecta may condense into dust. While such a model may not be realistic, it represents a case in which the predicted dust mass
at z$\sim$0  is comparable to the upper limit set by the most recent observations (e.g. Temi et al., 2004). Models with
lower dust production by SNeIa and/or stronger grain destruction do not meet such a
requirement. In the case of maximal production of dust (and minimal
destruction) the predicted [Fe/H] in the gas at z$\sim$0 can decrease by a factor of 3 with respect
to the standard predictions by the galactic wind models, hence reconciling the model with observations. 
In the less extreme and more realistic case presented here (Model 1ad, dashed line in the upper panel of Fig.~\ref{fig1}),
dust lowers the predicted [Fe/H] by $\sim$ 0.1 dex. This is not
enough to solve the theoretical Fe discrepancy, although it is likely that
the combined effect of dust and the other processes discussed in this section can solve the problem.\\


In the middle panel of Fig.~\ref{fig1}, we show that adopting the SNIa yields from
the Maeda et al. (2010) 2D explosion models (Model 3a, dot-dashed line) would reduce the predicted 
[Fe/H] by 0.1 dex. This change alone is small, but coupled with the dust would
suffice to reconcile model predictions to observations.
No significant differences with respect to the fiducial case are found when the WDD1 yields from Iwamoto et al. (1999)
are adopted.

Finally, let us examine different possibilities grouped under the broad cases \emph{4} and \emph{5}
(Fig.~\ref{fig1}, lower panel).
As we have seen in Sec.~\ref{thecode} (Fig.~\ref{fig0}),  different progenitors
change the SNIa rate. 
We first tested the empirical SNIa rate suggested by Mannucci et al. (2006).
In this case,  the total number of SNIa ever exploded and the
present-day SNIa rate are very close to our fiducial model 1a (Matteucci et al., 2006). 
{ This happens
because the t$>$ 3 Gyr evolution in the predicted
SNIa rates are very similar (c.f. Fig. 7 in Matteucci et al., 2006, Figs. 3-5 in Valiante et al., 2009). }
However, the distribution in time of the SNIa event 
changes in a way that affects the fraction of Fe ejected in the wind
as opposed to the Fe trapped in the hot ISM. The net effect is a slight decrease in the average SNIa
rate in the recent past (t$>$6Gyr) which leads to a reduction in the [Fe/H] abundance
by $< 0.1$ dex. Therefore, we do not discuss it further.
To make the models agree with observations, a more significant (a factor of 6-9) reduction
in the SNIa rate is required (e.g. Loewenstein \& Mathews, 1991, Renzini et al., 1993, Loewenstein
\& Davis, 2010). However, these solutions deserve more careful modelling by 
means of chemical evolution simulations and comparison with the latest 
observations.
We did this by reducing the constant $A$ in Eq.~\ref{snia} by a factor of 9.
The group \emph{4} of models differ only in the time at which the transition
from $A=0.09$ to $A=0.01$ occurs.
A very early occurrence (at $t=t_{GW}$, model 4a) is not the best solution, because 
the model is not able to maintain a wind. Therefore, even if the total number
of SNIa is less, their ejecta accumulate in the hot ISM for a longer time.
Model 4c works much better. 
A similar result can be achieved by adopting a completely different SNIa
scenario (DD, model 5a, dotted line in the lower panel of Fig.~\ref{fig1}),
at the expenses of a much lower [Fe/H] during the star forming phase.
While all the previous models reproduce the [Mg/Fe] ratio observed in the stars
of present-day ellipticals, model 5a exceeds the observed value because
of the strongly suppressed production of Fe during the star forming phase.
Also, such a model is not able to maintain a long lasting wind.

We interpret this results as a need for reducing the \emph{effective} SNIa rate
after the wind. However, a change caused by the nature of SNIa progenitors is
not necessarily the best physical explanation.
Perhaps, either differential winds or a differential (faster) cooling of SNIa ejecta 
is the solution.

The fiducial model predicts the [Fe/H] in the ISM to be a factor
of 2-3 higher than the average Fe abundance in stars (c.f. Table~\ref{setup}). Here we recall that
during the star forming phase, the abundance of a given
element in the gas is similar to that in the stellar population formed last
and certainly higher than the average abundance of that element in the stars. 
During the galactic wind phase, instead, most of the gas (and hence the metals) are
ejected from the galaxy and pollute the Intracluster/Intergalactic Medium (ICM/IGM).
The ISM in later phase is then heavily polluted by SNe Ia. 
Therefore, there is no reason to expect the Fe abundance in the hot ISM to be similar
to that in the stars. However, this is what is observed (e.g. Humphrey \& Buote, 2006).
The models (\emph{1,2,3 and 4}), that we introduced to address the \emph{theoretical} Fe discrepancy,
reduce the [Fe/H] ratio in the ISM without changing the average abundance in stars, hence
reconciling the models with the observations. Model 5a, on the other hand, strongly affects
the average stellar Fe abundance, thus it does not work in the right direction.
Here we mention that the Fe abundance inferred from stellar absorption lines is dependent
on the adopted synthetic simple stellar population\footnote{For instance, in the widely
used models by Thomas et al. (2003) [Fe/H] is not an independent quantity, 
but it is derived after the total metallicity and $\alpha$-enhancement
are measured, see the discussion in Pipino et al. (2006).} and sometimes
it is confused with the average metallicity ([Z/H]) in stars. Finally,
average abundances derived from Lick indices, while robust in relative sense (i.e. when comparing
two galaxies), may have systematics affecting their absolute values.
These factors render the comparison between the ratios in the stars to those in the
gas subject to systematics.

We conclude that models with either a realistic treatment of the dust or a variation of the yields
alone, are in disagreement with the observations. Their combined effect
is likely to solve the \emph{theoretical} Fe discrepancy.
On the other hand, a change in the \emph{effective} SNIa rate has a larger impact, even more than that
of a sudden dilution with pristine gas.
In passing, we note that a change in the IMF would not affect the conclusions of this
section, if the SNIa rate of such a model is tied to the observational present-day value.

\section{Abundance ratios in the ISM}
\label{abu}

Once  examined the temporal evolution of the [Fe/H] in the ISM of our models 
and understood how to fix the theoretical Fe discrepancy, we can have a look
at the consequences of these assumptions on the ISM abundance pattern.


Fig.~\ref{fig2} shows the predictions for the temporal evolution of the [$\alpha$/Fe] ratios as predicted
by our fiducial model 1a (solid lines). The shaded areas represent
the observed range of values in NGC 4472 from Loewenstein \& Davis (2010). { For simplicity
we use the values obtained with Suzaku as the mean values, whereas the scatter is giving
the 99.6\% confidence levels\footnote{We interpreted the 90\% observational uncertainties from the first row of Table 
1 in Loewenstein \& Davis (2010) as typical $\sim$ 1.6 $\sigma$  
errors . We then propagated 
to the abundance ratios with the standard formulae. We then show the so-calculated 3 $\sigma$ (99.6\% regions) error 
in the plots.}.} { The [Fe/H] evolution from Fig.~\ref{fig1} is also reported for comparison to the
specific Fe abundance measured in NGC 4472}.
We took the recent observations for this single galaxy as a proxy for the typical pattern (see also
Kim \& Fabbiano, 2004, Humphrey \& Buote, 2006, Ji et al., 2009). It is important to point out that
when the instrument-to-instrument and galaxy-to-galaxy differences are taken into account the observed range increases.
We can clearly see that the { higher than observed} Fe abundance predicted by model 1a implies low  [$\alpha$/Fe] ratios,
at variance with observations. { Models 1afd, 3a and 4c, instead, make predictions that are
less than 0.1 dex away from the observed range in the specific case of NGC 4472. Such a discrepancy
may stem from the fact that the model is not explicitly tailored to reproduce this galaxy and there
might be small differences between the observed and predicted total gas (and H) content which  may
compensate for it. Abundance ratios, instead, do not have such a problem. }
Here we recall that, when SNIa ejecta dominate, Si and Ca are expected to be more enhanced than Mg and O, because
the former elements are produced in a non negligible way by SNIa.
Mg and O, instead, mostly come from the hydrostatic burnings in massive stars, therefore their 
abundances should evolve in lock-step.
Indeed, model 1a predicts [Mg/O]$\sim$0, whereas observations hint
for [Mg/O]$\sim$0.3, namely the O should be under-produced (or the Mg over-produced) by a factor of 2. 
Therefore we are interested in solutions to the theoretical Fe discrepancy put forward in Sec.~\ref{solutions} which lead to specific changes
in the predicted abundance ratios and could mitigate the disagreement with observations. 
A simple dilution with pristine gas, instead, would not affect the abundance ratios.

The top row of Fig.~\ref{fig2} shows the results for the models with dust compared to the fiducial case.
As far as the dust is concerned, it is worth noting that Fe and O have different condensation temperatures, being Fe a ``refractory''
element, hence more easily prone to dust depletion than O. Mg, instead, has nearly the same
condensation temperature of Fe (Lodders et al, 2003).
Therefore, if dust is involved with the canonical prescriptions by Pipino et al. (2011), 
the predicted [O/Fe] ratio in the gas increases (model 1ad), hence improving the agreement with 
observations. On the other hand, the increase (dashed lines in the top row) in the [O/Mg] worsen the departure
from observations.
At the same time, the predicted [Mg/Fe] ratio remains basically the same, hence still below the observed value.
A preferential condensation of Fe with respect to Mg seems to be suggested also by high-redshift
observations of Lyman Break Galaxies (Pipino et al., 2011), but currently no theoretical
model for the dust formation can explain it.
Ca and Si also have condensation temperatures similar to Fe. For the Si, the same
hints of a differential dust depletion are inferred from high-redshift objects (Vladilo, 2002, Calura et al., 2003,
Pipino et al., 2011). 
On the other hand, if Fe preferentially condense (model 1afd, dot-dot-dashed line) the
theoretical predictions make the [Mg/Fe] and [O/Fe] ratios in the hot ISM to increase in lock step.
Therefore we would still predict [Mg/O]$\sim$0, at variance with observations\footnote{Assuming that
the high [Mg/O] ratio in the ISM is not an instrumental artifact. Otherwise many more models
would fit the observed abundance pattern.}.
This means that the dust alone cannot solve the Fe discrepancy and the abundance pattern at the same time.

Let us now focus on the middle row of Fig.~\ref{fig2}, namely on the effects of SNIa nucleosynthesis.
Model 1c predictions (namely the case with \emph{5xMg} in SNIa yields- dotted lines) improve upon the fiducial case. The predicted
[Mg/Fe] is still well below the observed value, and the prediction does not gain more than 0.1 dex by adding
Fe-only dust (model 2cfd, not shown), however the predicted [O/Mg] ratio is in marginal
agreement with the observations. Of course, these changes do not affect the predictions regarding [Ca/Fe] and [Si/Fe].

Both the reduced Fe production and the increase in the Mg yield in the C-DDT SNIa models make
model 3a (dot-dashed lines) very close in reproducing [Mg/Fe], [O/Mg] and [Si/Fe] at the same
time. { We stress that, albeit the \emph{agreement} is still only  at a 3$\sigma$ level, it may become
smaller when the systematic uncertainties (instrument-to-instrument) variations are taken into account. 
Similarly, taking into account metallicity-dependent yields and/or a higher cutoff for the IMF
will improve the agreement (see discussion below).} 
In practice, the yields from Maeda et al. (2010) go in the direction suggested by 
the empirical yields by Fran\c cois et al. (2004).
When such a choice for the SNIa yield is coupled to, e.g., only Fe dust and to some dilution,
it may work for reproducing a larger set of observables. Unfortunately, the predictions
for the [Ca/Fe] do not improve with respect to the fiducial case, hence differing
from the observation by $\sim$0.5 dex.
When adopting the WDD1 SNIa yields (model 2a, dashed line), the predictions for the [Ca/Fe]
ratio are in good agreement with observations. The [Si/Fe] ratio is recovered, 
whereas predictions for the [Mg/Fe] and the [O/Mg] ratios basically coincide with those
of the fiducial model. 
As noted by Loewenstein \& Davis (2010), SNIa yields which best reproduce the Ca abundance
fail in reproducing the Ni one (Fig.~\ref{fig3}) and vice-versa.
Our results agree with those found by Loewenstein \& Davis (2010). Also,
similar claims have been made by other authors (e.g. Finoguenov et al., 2002),
but their quantitative conclusions are less certain, being based on a direct comparison to the stellar yields,
without taking into account the recycling.
Therefore, we do not attempt a direct comparison.
{ Furthermore, we wish to stress that, while Mg, Si and  O have been measured by different authors, the Ca abundance
is less constraining. Therefore, the above attempt to find a model that also fits the [Ca/Fe] ratio should be
regarded just as an exercise, until more robust observational estimates become available for a larger number of galaxies.}
Similar results apply to the predictions regarding [Ne/Fe], not shown here.

We briefly mention that the [$\alpha$/Fe] ratio in SNII ejecta increases with decreasing metallicity
(Kobayashi et al., 2006). As a consequence, if we used the model 3b instead of 3a we would increase each
[$\alpha$/Fe] ratio by $\sim 0.2$ dex at most\footnote{The reader should remind that the material
produced by SNII is locked-up in low-mass long-living stars and then ejected in the ISM where
it mixes with SNIa ejecta.}, with the largest effect for [Ca/Fe]. Of course, only a fraction of the stars will be formed at such low
metallicities. Therefore, the true effect of the metallicity dependent yields by Kobayashi et al. (2006)
will be probably $\sim$ 0.1 dex.
{ It is worth noting that the [$\alpha$/Fe] ratios in SNII ejecta increase with stellar mass (c.f. Fig. 2 in Kobayashi
et al., 2006). If we extrapolate this trend and adopt 70 M$_{\odot}$ as a the upper mass limit for the IMF in the models,
the predicted [O,Mg/Fe] ratios in the hot ISM may increase by $\sim$0.1 dex, whereas [Fe/H] is not
affected. We avoid this choice
in our fiducial models in order not to introduce the additional uncertainty linked to the extrapolation
of behaviors which are (in some cases) far from linear. Developments in this sense would require a more
extended grid of SNII yields. Nonetheless, a higher mass limit in the IMF would help in improving the agreement
between the most promising models and the observations as far as [O/Fe] and [Mg/Fe] are concerned. Such a choice would
not change, instead, the predicted [O/Mg] ratio, since in the Kobayashi et al. (2006) calculations, their
yields exhibit a similar increase with the progenitor mass.}
On the other hand, stars suffer from mass loss that increases with increasing metallicity\footnote{In the gas
out of which the star formed}. The larger the mass loss, the smaller the He and C cores and, hence,
the O production. Therefore, when the effect of the mass loss is accounted for, the predicted 
[O/Mg] ratio in the gas decreases with metallicity during the star forming phase. Such an effect has been observed, 
and explained by means of mass loss in massive stars,
in the most metal rich stars of the Milky Way disc (McWilliam et al., 2008) as well as in the Galactic Bulge (Cescutti et al., 2009).
If this finding is true, we may expect that some of the gas locked in low-mass metal-rich stars
is indeed O-depleted. When these stars expel their envelopes in the planetary nebula phase they may
slightly lower the O/Mg ratio in the ISM.

Finally, let us discuss the effects of changing either the SNIa explosion rate or their progenitors
(lower panel in Fig.~\ref{fig2}). As expected, a suppression of the SNIa rate can easily
bring several [$\alpha$/Fe] ratios well in agreement with the observations. Nonetheless, these models cannot
explain the observed [O/Mg] ratio. Nor do they give a reasonable fit in the case
of [Ca/Fe].
However, the major problem with this scenario is represented by the observed SNIa
rate.

\section{Discussion}

In general, the observed abundance ratios exhibit
a pattern in which some of them (Ca, Si) are solar to over-solar, other are solar (Mg), whereas the O is depleted
with respect to Fe.
This is at variance with what happens in the stars (c.f. Table~\ref{setup}), where Mg (and probably
O) are enhanced, whereas Ca/Fe is solar, and at odds  with the behavior of the [Fe/H] in the ISM,
which tracks the average Fe abundance in the stars.
Such a clear asymmetry (Renzini et al., 1993) between
$\alpha$-enhanced stars, and Fe-enhanced hot ISM is well understood in terms
of SNeII which dominate the early phases in the evolution, when the stars 
are formed,
and SNeIa which dominate the late time ISM abundance pattern.
However, the situation is not so clear-cut and not all the $\alpha$-elements have the same
degree of enhancement (depletion) in the stars (hot ISM). The yields for each
given element, as well as the recycling by low-mass single stars do play a role
is shaping the abundance pattern. 
Matteucci \& Chiappini (2005),  and more recently Loewenstein \& Davis (2010) 
stressed the fact that in order to reproduce the abundance ratios
of many elements it is important to have an $\alpha$-enhanced gas ejected (recycled) by single
low-mass stars. That is, the abundance ratios cannot be interpreted
by comparing them directly to the predicted SNIa/SNII yield ratio pattern.
This is true even if they involve $\alpha$-elements, namely
elements that one naively expects to be produced on the same 
(negligible) timescale.

Therefore, we argue that the use of abundance ratios provides new insight
into the chemical evolution of elliptical galaxies.
Either some dilution/differential ejection or a Fe-only dust models help 
in achieving a better agreement with observables once a suitable modification to the SNIa yield with respect to
the standard W7 model is adopted, being the 2D explosion model C-DDT the best candidate for reproducing
both the ISM [Fe/H] and the [$\alpha$/Fe] ratios in the specific case of NGC 4472.
{ For this specific galaxy, the average stellar [Fe/H] and [Mg/Fe] ratios
are $-0.01(\pm 0.2)$ dex and $0.26(\pm 0.03)$ dex, respectively (Humphrey \& Buote, 2006, Thomas et al., 2005), hence
all models 1,2,3 and 4 are consistent within 1 $\sigma$ as far as the Fe abundance is concerned. These models are consistent
with the observationally inferred stellar [Mg/Fe] ratio within 2 $\sigma$\footnote{ With the additional caveat
that the stellar value values are robust only in relative sense.}, with
the best match given by model 2. 
As already discussed, model 5 is at odds with the stellar abundance pattern.
An interesting follow-up study would be to (re-)derive the stellar abundances for a more extended set of elements
by means of newer stellar models (e.g. Graves \& Schiavon, 2008, Thomas et al., 2010) in a sample
of galaxies which have homogeneous X-ray observations, in the spirit of Humphrey \& Buote (2006).}

Finally, we spend a few words on alternative scenarios not explored here.
Hierarchical models (e.g. Kawata \& Gibson, 2003, Bekki, 1998, Cox et al., 2006), in general
do not give satisfactory answers for the opposite
reason, namely they predict a too low Fe abundance in the ISM. We note here that some of these
models are built on simulations in which equal mass disks merge and that they have well known difficulties in explaining the chemical properties 
of the stars in elliptical galaxies (Ostriker, 1980, Thomas 1999, Pipino \& Matteucci, 2006, Pipino et al., 2009b).
Other models have a galaxy formation process that is more similar to the monolithic one (e.g. Kawata \& Gibson, 2003),
however they are characterized by strongly under-solar
$[\alpha/Fe]$ ratios.

A larger
contribution from SNeII may be invoked to explain the observed abundance 
pattern.
This can be easily done by requiring a flatter (e.g. x=0.95) IMF than the Salpeter one.
In this particular case, the [O/Mg] ratio will not change, hence still requiring a fix in the SNIa nucleosynthesis.
Moreover, for a given set of SNIa progenitors and requiring that the model
still reproduces the present-day SNIa rate, the predicted [Fe/H] will be still
higher than the observed one.
Finally, if a flat IMF is kept for the entire galactic evolution, the larger amount
of metals promptly released by SNeII, will make the stars of the galaxies too metal-rich
at variance with observations (Pipino \& Matteucci, 2004) and will not
help in reproducing the observed trend of the stellar [Mg/Fe] with galactic mass (Nagashima et al., 2005).
If the flat IMF is confined only to the very early phases in which the so-called
Pop III (metal-free) stars have been formed, such a regime is so short in time that
very few stars are created, compared to the total, therefore their impact should be minimal (e.g.
Matteucci \& Pipino, 2005).

\begin{figure}
\begin{center}
\includegraphics[height=8cm,width=8cm]{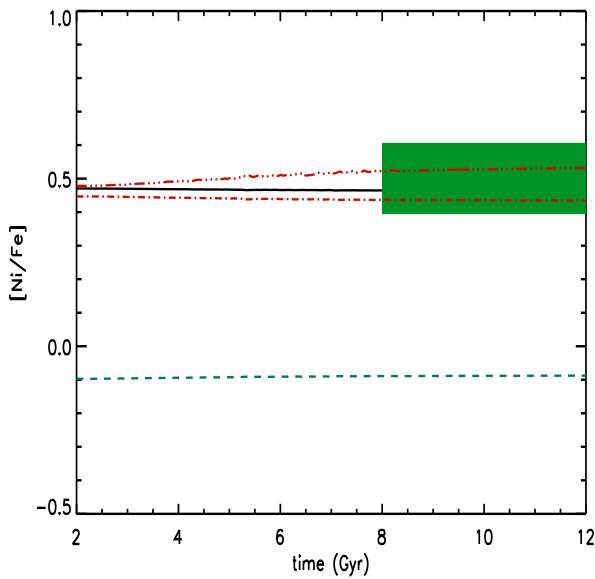}
\caption[]{[Ni/Fe] abundance ratio versus time. The shaded areas represent
the observed range of values and the 99\% uncertainty. 
Fiducial model 1a (solid) compared to models 1c (dotted), 2a (dashed), 3a (dot-dashed), respectively.
\label{fig3}}
\end{center}
\end{figure}

\section{Conclusions}
\label{concl}

The galactic wind models well reproduce the abundance pattern in stars
of elliptical galaxies,
whereas they predict too much Fe to be released in the hot ISM by SNeIa.
Several solutions have been put forward, and they basically tell us that
what happens after the galactic wind phase is a rather complicate mixture
of processes which, perhaps, cannot be entirely explained with 1D symmetrical simulations
such as chemical evolution models.
We reviewed several of these solutions suggested to explain the Fe
abundance and the abundance ratios pattern in the hot ISM
of elliptical galaxies.  We created a comprehensive grid of models
to test them by using a full chemical evolution scheme.

As far as the Fe abundance is concerned, we can conclude that in models with either a realistic treatment of 
the dust or a variation of the yields
alone,  the disagreement with the observations is strongly reduced. Their combined effect
solves the \emph{theoretical} Fe discrepancy.
On the other hand, a change in the \emph{effective} SNIa rate (after the wind) has a greater impact, even larger than that
of a sudden dilution with pristine gas.
A change caused by the nature of SNIa progenitors is
not necessarily the best physical explanation.
Perhaps, either differential winds or a differential (faster) cooling of SNIa ejecta 
which mimic a reduction in the effective SNIa rate are a better physical explanation.
We note that a change in the IMF would not affect the conclusions,
 if the SNIa rate of such a model is tied to the observational present-day value.
However, these models alone cannot explain the observed abundance ratios and the differences
between the single $\alpha$ elements.

{We consider the 2D explosion model C-DDT as the most promising candidate for reproducing
[Fe/H] and the [$\alpha$/Fe] ratios in both stars and ISM}. 
Only the Ca seems at odds, { although it should be measured both in a larger number of
galaxies  and with a smaller uncertainty before reaching firm conclusions.}
Therefore, we are more optimistic than Loewenstein \& Davis (2010), who note that
there seems not be any combination of nucleosynthetic yields from both SNII and SNIa
that produces the observed abundance pattern in the hot ISM.
Indeed, we think that improvement in SNIa explosion calculations
may offer a final solution in the near future.
Fe-only dust may help 
in achieving a better agreement with the observables.
Unfortunately, we cannot derive the O, Si, Ni abundances in integrated
stellar spectra, thus some suggested solutions are still unconstrained.

\section{Acknowledgments}
We warmly thank the referee M. Loewenstein for comments which improved the quality
of the paper.
Useful comments by D.-W. Kim, C. Kobayashi 
are also acknowledged.
AP received support from Italian Space Agency
through contract ASI-INAF I/016/07/0 
F.M acknowledges financial support from PRIN2007 MIUR (Italian Ministry of Scientific Research) 
project Prot. no. 2007JJC53X-001.

\end{document}